\documentstyle[epsfig]{mn}
\begin{document}
\def\simlt{\mathrel{\rlap{\lower 3pt\hbox{$\sim$}}\raise 2.0pt\hbox{$<$}}}
\def\simgt{\mathrel{\rlap{\lower 3pt\hbox{$\sim$}} \raise 2.0pt\hbox{$>$}}}
\def\Msun{M_{\odot}}

\title[Where are the sources of the NIRB?]{Where are the sources of the near infrared background?}
\author[Salvaterra \& Ferrara]
{Ruben Salvaterra$^1$ \& Andrea Ferrara$^2$ \\
$^1$Dipartimento di Fisica e Matematica, Universit\'a dell'Insubria, Via Valleggio 11, 22100 Como, Italy\\
$^2$SISSA/International School for Advanced Studies, Via Beirut 4, 34100 Trieste, Italy}

\maketitle \vspace {7cm}

\begin{abstract}
The observed near infrared background excess over light from known galaxies is commonly ascribed to redshifted radiation from early, very massive, PopIII stars. We show here that this interpretation must be discarded as it largely overpredicts the number of $J$-dropouts and Ly$\alpha$ emitters in ultra deep field searches. Independently of the detailed physics of Ly$\alpha$ line emission, $J$-dropouts limit the background excess fraction due to PopIII sources to be (at best) $\le 1/24$. As alternative explanations can either be rejected (e.g. miniquasars, decaying neutrinos) or appear unlikely (zodiacal light), whereas the reality of the excess is supported by the interpretation of the angular fluctuations, the origin of this component remains very puzzling. We briefly discuss possible hints to solve the problem.   
\end{abstract}


\section{Introduction}

Recent measurements of the Near Infrared Background (NIRB) 
(see Hauser \& Dwek 2001 for a review) have shown an intensity excess with 
respect to observed light from galaxies in deep field surveys (Madau \& 
Pozzetti 2000; Totani et al. 2001). 
The discrepancy is maximal at $1.4\;\mu$m, corresponding 
to $17-48$~nW~m$^{-2}$~sr$^{-1}$ (about 2-5 times the known galaxy 
contribution)(Matsumoto et al. 2005); 
significative discrepancies are found also at longer wavelengths. The large 
uncertainty on the amplitude of the excess is due to the  
subtraction of the zodiacal light (i.e. sunlight scattered by
the interplanetary dust) contribution. Two different models for the 
zodiacal light have been so far proposed, which predict a higher (Wright 1998),
or a lower (Kelsall et al. 1998) contribution of this component to he NIRB.
It has to be noted, though, that the excess cannot be completely 
explained by a combination of galaxy and zodiacal light alone,
unless current zodiacal light models are seriously incorrect.

Estimates based on theoretical models suggest that the most likely origin of 
such excess is redshifted light from early, very massive ($M>100\Msun$) 
PopIII stars (Santos, Bromm \& Kamionkowski 2002; Salvaterra \& Ferrara 2003 [SF]; Fernandez \& Komatsu 2005) 
if these form efficiently down to $z\approx 8$. The same stars can also account 
for the observed small scale angular fluctuations detected in the same bands 
(Magliocchetti, Salvaterra \& Ferrara 2003; Kashlinsky et al. 2004; Cooray
et al. 2004).
In this paper, we review this conclusion in the light of the new {\it Spitzer} satellite 
and ground-based ultra deep searches for high redshift galaxies.
Throughout the paper we use the AB magnitude system\footnote{AB magnitudes are defined
as AB$=-2.5\log (F_{\nu_{0}}) -48.6$, where $F_{\nu_{0}}$ is the spectral energy density
within a given passband in units of erg s$^{-1}$ cm$^{-2}$ Hz$^{-1}$}.

\section{Basic Equations} 

The mean specific background intensity,
$J(\nu_{0},z_{0})$, at redshift 
$z_{0}$ observed at frequency $\nu_0$, 
produced by a population of sources (in our case, the PopIII stellar clusters)
characterized by a comoving emissivity 
$\epsilon_\nu(z)$, can be written as
\begin{equation}\label{eq:I}
J(\nu_{0},z_{0})= \frac{(1+z_0)^3}{4\pi}\int^{\infty}_{z_{0}} \epsilon_\nu(z)
\mbox{e}^{-\tau_{eff}(\nu_0,z_0,z)} \frac{dl}{dz}dz,
\end{equation}
where $\nu=\nu_0(1+z)/(1+z_0)$, $dl/dz$ is the proper line element, 
$\tau_{eff}$ is the IGM effective optical depth at $\nu_0$ between
redshift $z_0$ and $z$ (see Sect. 2.2 of SF for a full description of 
the IGM modelling). 

The source term $\epsilon_\nu$ can be written, in general, as a convolution of 
the light curve of the stellar cluster $l_{\nu}(t)$ with the source formation rate 
(per unit comoving volume). Under the approximation that the source formation 
rate is constant over the source lifetime $\tau_{lf}$, and considering the 
average specific luminosity per unit mass, $\bar l_{\nu}$, over such timescale,
the emissivity is given by

\begin{equation}\label{eq:emis}
\epsilon_\nu(z)\simeq \bar   l_{\nu} \tau_{lf} f_\star \frac{\Omega_b}{\Omega_m}
\frac{d}{dt} \int^{\infty}_{M_{min}(z)}  \frac{d n}{d M_h}(M_h,z) M_h  dM_h,
\end{equation}
where $n(M_h,z)$ is the comoving number density of halos of mass $M_h$
at redshift $z$ given by Press \& Schechter (1974). The integral
represents the collapsed mass per comoving volume contained in dark matter halos 
with mass above $M_{min}(z)$. We consider two 
different threshold masses depending on the relevant cooling agents:    
(i) atomic hydrogen (i.e. $T_{vir}>10^4$ K); and (ii) 
molecular hydrogen as computed by Fuller \& Couchman 
(2000; hereafter FC).
The time derivative converts the collapse mass per 
comoving volume into a formation rate and the pre-factor 
$f_\star (\Omega_b/\Omega_m)$\footnote{We adopt the `concordance' model values for the cosmological parameters: $h=0.7$, 
$\Omega_m=0.3$, $\Omega_{\Lambda}=0.7$, $\Omega_b=0.038$, $\sigma_8=0.9$,
and $\Gamma=0.21$.} converts from total mass to stellar mass; 
$f_\star$ is the PopIII star formation efficiency.

As for PopIII stars properties, we adopt the light curves of massive metal-free 
stars (Schaerer 2002), including nebular emission and Ly$\alpha$ emission line 
(Loeb \& Rybicki 1999; for a detailed discussion, see SF). 
For simplicity, we consider a $\delta-$function IMF peaked 
at 300 $\Msun$, since PopIII specific spectra above 
$M_\star>100\Msun$ are essentially identical 
(Bromm, Kudritzki \& Loeb 2001). Stellar lifetimes are also very similar
for very massive stars, so we adopt a single value $\tau_{lf}=2$~Myr.

\begin{figure}
\center{{
\epsfig{figure=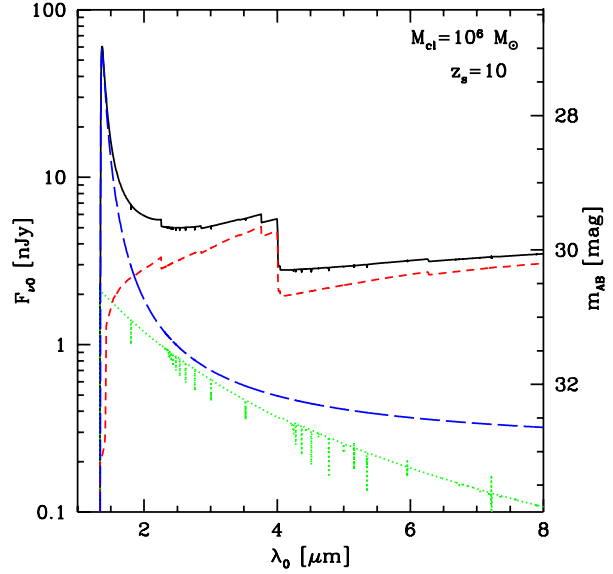,height=8cm} }}
\caption{\label{fig:flux} Observed flux from a $M_{cl}=10^6\;\Msun$ PopIII stellar
cluster 
at $z_s=10$. The lines show the total flux (solid), and the stellar 
(dotted), nebular (dashed), and Ly$\alpha$ line (long-dashed) contributions.}
\end{figure}

The average flux $F_{\nu_0}(z,M_h)$ from PopIII stellar cluster of mass $M_{cl}$
at redshift $z$ is given by

\begin{equation}\label{eq:flux}
F_{\nu_0}=\frac{M_{cl}}{4\pi \Delta\nu_0\, d_L(z)^2} \int^{\nu_{max}}_{\nu_{min}} \bar l_\nu \mbox{e}^{-\tau_{eff}(\nu_0,z_0,z)} d\nu,
\end{equation}

\noindent
where $d_L(z)$ is the luminosity distance, $\Delta\nu_0$ is the instrumental bandwidth, 
and $\nu_{min}, \nu_{max}$ are the restframe frequencies corresponding to the observed ones. 
In Fig. \ref{fig:flux} we show the expected flux from a PopIII stellar cluster 
 with $M_{cl}=10^6\;\Msun$ at redshift $z_s=10$. It is interesting to 
note that such an object could be luminous enough to be detected at the NICMOS 
$H_{160}=28$ limiting magnitude.

\bigskip

Eqs. (\ref{eq:I})-(\ref{eq:emis}) allow us to compute the diffuse NIRB 
background due to PopIII stars and compare it with the data. 
In order to determine the minimum requirements for the sources responsible for the 
NIR light, we consider the maximal contribution of both ``normal'' deep field galaxies 
(Madau \& Pozzetti 2000; Totani et al. 2001) and zodiacal light (Wright 1998). 
We then determine the minimum star formation efficiency, $f_\star^m$, necessary to 
fit the NIRB spectrum (additionally, we assume that PopIII stars do not
form below $z_{end}\approx 8$, as implied by the background spectrum $J$-band peak). 
We use here the data obtained by DIRBE and the recent determination
by IRST (Matsumoto et al. 2005) that provide important information on the shape
of the NIRB in the 1.4-4 $\mu$m wavelength range, rescaled using the Wright 
(1998) zodiacal light model subtraction.
We will refer to this solution as the {\it minimal} NIRB model, for which 
$f_\star^m=0.8$ (0.4) for H-cooling (H$_2$-cooling) halos. Note that this value is
larger than derived by SF, due to their simplifying assumption that $\tau_{lf}$ is 
of the order of the Hubble time. Finally, for the given value of $f_\star^m$, we compute  
the surface density of PopIII stellar clusters required to fit the {\it minimal} model   
in the flux range $F_{\nu_0}$ and $F_{\nu_0}+dF_{\nu_0}$ as

\begin{equation}
\frac{dN}{d\Omega dF_{\nu_0}}(F_{\nu_0},z_{\rm end})=\int^{\infty}_{z_{\rm end}}
\left( \frac{dV_c}{dz d\Omega} \right) n_c(z,F_{\nu_0}) dz,
\end{equation}

\noindent
where $dV_c/dz d\Omega$ is the comoving volume element per unit redshift per
unit solid angle, $n_c(z,F_{\nu_0})$ is the comoving number of objects at
redshift $z$ with observed flux in $[F_{\nu_0}, F_{\nu_0} + dF_{\nu_0}]$, given by

\begin{equation}
n_c(z,F_{\nu_0})\simeq \frac{dM_h}{dF_{\nu_0}}(z,F_{\nu_0}) \tau_{lf} 
\frac{d^2 n}{dM_h dt}(M_h,z).
\end{equation}

\noindent

\section{OBSERVATIONAL CONSTRAINTS}

\subsection{Spitzer number counts}

\begin{figure}
\center{{
\epsfig{figure=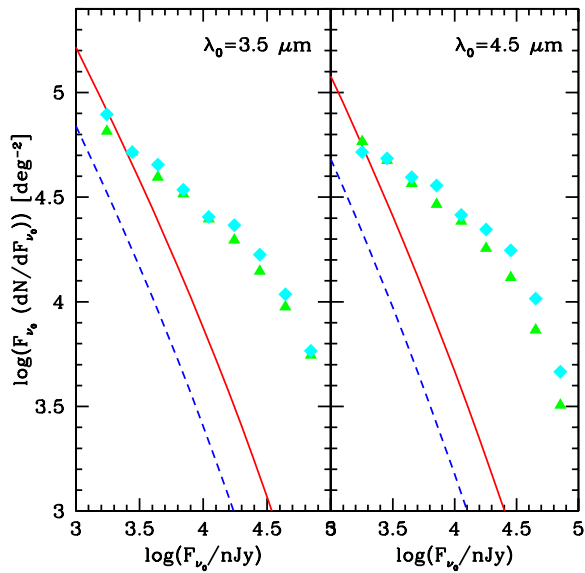,height=8cm} }}
\caption{\label{fig:SPITZER} Expected number counts for PopIII sources fitting the NIRB excess. 
The solid (dashed) line refers to H-cooling (H$_2$-cooling) halos.
{\it Left panel:} $\lambda=3.5\;\mu$m, {\it right panel:}
$\lambda=4.5\;\mu$m. Data (Fazio et al. 2004b) points are from the {\it Spitzer} 
QSO HS1700+6416 field (diamonds) and the Extended Groth Strip (triangles).
}
\end{figure}

Ground-based mid-infrared (MIR) galaxy counts are limited by the
high thermal background emission and narrow atmospheric transmission window. 
The InfraRed Array Camera (IRAC) instrument on the {\it Spitzer Space Telescope} provided the first
source counts in the wavelength range $3<\lambda_0<10\;\mu$m. Thanks to the low
background available in space, IRAC can reach faint limiting fluxes
in modest exposure times. Fazio et al. (2004b) reported the combined data from
three survey fields of complementary area and depth at four IRAC wavelengths,
3.6, 4.5, 5.8, and 8.0 $\mu$m, respectively. The 3.5 and 4.5 $\mu$m images are affected by
source confusion at their faint limits; for this reason, Fazio et al. (2004b) choose
to truncate the source counts at the relatively bright flux value of 
$1.7\mu$Jy\footnote{In the original paper of Fazio et al. (2004b) 
data are given  in instrumental magnitude relative to Vega. Throughout the 
paper we convert these into fluxes for clarity. The flux density of Vega in the 
3.5 $\mu$m (4.5 $\mu$m) IRAC bandpass is 277.5 (179.5) Jy (Fazio et al. 
2004a).}. After star subtraction and completeness correction, Fazio et al. 
(2004b) provided extragalactic source counts for the three fields and the four 
IRAC wavelengths. The data for the 3.5 and 4.5 $\mu$m images are shown in Fig. 
\ref{fig:SPITZER} for the two deepest fields, around the  
QSO HS1700+6416 and  in the Extended Groth
Strip. Data the (widest) field in the Bootes region are not displayed due     
to their higher flux limits.

\begin{figure}
\center{{
\epsfig{figure=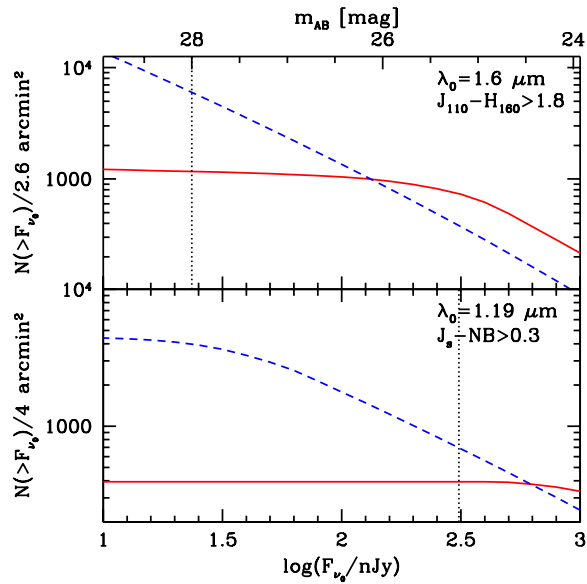,height=8cm} }}
\caption{\label{fig:NICMOS} {\it Top panel}: number of $z\sim 10$
sources predicted by the {\it minimal} NIRB model. The selection criterion
$J_{110}-H_{160}>1.8$ has been applied and  
a FOV of 2.6 arcmin$^2$ has been assumed for comparison with NICMOS UDFs. The dotted 
line shows the conservative flux limit for the NICMOS survey of Bouwens et al. (2005).
{\it Bottom panel}: number of sources with strong Ly$\alpha$ line 
emission predicted by the {\it minimal} NIRB model. 
The selection criterion $J_s-NB\geq 0.3$ has been applied and a FOV      
view of 4 arcmin$^2$ has been assumed. The dotted 
line shows the flux limit of the VLT/ISAAC survey of Willis \& Courbin (2005).
In both panels the solid (dashed) line refers to H-cooling (H$_2$-cooling) halos. 
Counts flatten below a certain flux as the threshold mass is reached.}
\end{figure}

\subsection{High redshift galaxy searches}

The detection of galaxies at the highest redshifts is challenging because
of the redshifting of UV light into the infrared and the well-known limitations
of current IR instruments. In spite of this difficulties, the search for high 
redshift galaxies, based on the dropout technique, has provided the
first tentative detection of galaxies at redshifts beyond $z\approx 7$ (Bouwens et al. 
2004), stimulating a re-analysis of the small amount of deep IR data.

Bouwens et al. (2005) looked at the prevalence of galaxies at $z\approx 8-12$ 
by applying the dropout technique to the wide variety of deep
F110W- and F160W-band (hereafter $J_{110}$ and $H_{160}$, respectively) fields
that have been imaged with the Near Infrared Camera and Multi-Object 
Spectrometer (NICMOS). The principal data set is constituted by two 1.3 arcmin$^2$ deep NICMOS parallels
taken with the Advanced Camera for Surveys (ACS) Hubble Ultra Deep Field 
(UDF; each has about $160$ orbits data).
Complementary shallower fields possessing similar $J_{110}+H_{160}$ imaging 
are analysed as well. The 5$\sigma$ limiting magnitude for the UDFs        
is  $\approx 28.6$ in $J_{110}$ and $\approx 28.5$ in $H_{160}$ 
($0^{\prime\prime}.6$ aperture). Since these limiting magnitudes are 
reached only in the deepest regions of the two NICMOS UDFs, we adopted the 
conservative limiting magnitude $H_{160}=28$ (Bouwens, private communication). 
The primary selection criterion for high-$z$
sources is $J_{110}-H_{160}>1.8$. Using this criterion, Bouwens et al. (2005) found
eleven sources. Eight of these are ruled out as credible $z\simeq 10$
sources, either as a result of detection ($>2\sigma$) blueward of $J_{110}$
or because of their colors redward of the break ($H_{160}-K \approx 1.5$). The nature
of the three remaining sources could not be assessed   from the data, but
this number appears consistent with the expected contamination from 
low-redshift interlopers. Hence, Bouwens et al. (2005) concluded that
the actual number of $z\simeq 10$ sources in the NICMOS parallel fields 
must be three or fewer.

Willis \& Courbin (2005) have presented a deep, narrow $J$-band search (called ZEN) for protogalactic Ly$\alpha$ emission at redshift $z\approx 9$, combining a 
very deep VLT/ISAAC image of the Hubble Deep Field South using a narrow-band (hereafter NB) 
filter centered at 1.187 $\mu$m, with existing deep, broad-band images covering optical and 
NIR wavelengths. They searched for $z\approx 9$ Ly$\alpha$-emitting
galaxies displaying a significant narrow-band excess relative to the
$J_s$ band (i.e. $J_s-NB\geq 0.3$) that are undetected at optical wavelengths. 
They identified no sources, matching these requirements, up to a NB magnitude 
limit of 25.2 in a field of $\sim 4$ arcmin$^2$.

\section{Implications for NIRB models}

Having fixed the {\it minimal} NIRB model (see Sec. 2), we ask what are
the predictions of such model in terms of: (i)  number counts in the $3.5$ and $4.5\;\mu$m {\it Spitzer}
bands,  (ii) number of $J$-dropouts that can be detected with NICMOS, and  (iii) number of strong Ly$\alpha$-emitting galaxies identifications.

The main results are shown in Figs. \ref{fig:SPITZER}-\ref{fig:NICMOS}
for the two cases of H- or H$_2$-cooling halos. 
In Fig. \ref{fig:SPITZER} we plot the predicted differential number counts
compared to {\it Spitzer} data; Fig. \ref{fig:NICMOS} shows the total number of 
$J$-dropouts expected in the NICMOS ultra deep observations (top panel)
and in the ZEN search (bottom panel).

If the NIRB excess is to be ascribed to PopIII clusters, we find that even for the {\it minimal}
NIRB model a substantial fraction of the {\it Spitzer} number counts is  provided by such 
galaxies, particularly at the faintest fluxes. 
For $T_{vir}>10^4$ K halos, almost all the (3.5 $\mu$m and 4.5 $\mu$m) 
IRAC counts at the flux limit of the survey should correspond to PopIII galaxies,
whereas for H$_2$ cooling halos $\approx 50$\% of the observed sources 
at $1.7\;\mu$Jy should lie at $z>8$. Even at fluxes of $4.4\;\mu$Jy, where
confusion noise and completeness problems are less important, $\approx $60\% 
(20\%) of the detected {\it Spitzer} sources should be PopIII galaxies in H-cooling 
(H$_2$ cooling) halos.
 
The situation becomes even more serious if we consider the predictions for the number of
$z\approx 10$ PopIII galaxies detectable in the NICMOS UDFs.       
Adopting the same selection criterion $J_{110}-H_{160}>1.8$
as in Bouwens et al. (2005), the model predicts that thousands of these sources should
be detected. More precisely, 1165 (5634) $J$-dropouts should be observed for H-cooling 
(H$_2$ cooling) halos below the conservative magnitude limit $H_{160}=28$. These numbers are at odd    
with the mere 3 (tentative) detections reported by Bouwens et al. (2005).

The $J$-band peak in the NIRB excess has been interpreted (SF; Santos et al. 2002) as
due to strong asymmetric Ly$\alpha$ emission line from PopIII clusters at
$z\approx 9$. Such sources can be identified in a narrow $J$-band survey
from a significant narrow-band ($NB$) excess relative to the flux 
observed in a broader $J$-band. In the bottom panel of Fig. \ref{fig:NICMOS}
we show that, by applying the criterion $J_s-NB\geq 0.3$, almost 394 (683)  
H-cooling (H$_2$-cooling) sources should be identified up to a magnitude limit $NB=25.2$ in a 
$4$ arcmin$^2$ field for the {\it minimal} NIRB model.
Instead, the ZEN search implementing the same selection criterion has 
yielded no identification. 

Can we reconcile the NIRB-PopIII interpretation with the constrain given by
the currently available observational data ?         
The {\it Spitzer} number counts can be reduced by           
quenching the formation of PopIII stars in larger halos as 
a result of the transition to a PopII mode driven by chemical feedback 
(Schneider et al. 2002, 2003, 2005). 
In this case the {\it minimal} NIRB model is reproduced if PopIII star formation occurs in all halos with masses in the range $M_{min}(\mbox{FC}) < M_h \simlt 10^9 \Msun$ with $f_\star^m=0.8$. Chemical feedback would drastically reduce the 
number of PopIII IRAC galaxies without reducing the NIRB intensity, 
the latter being mainly sustained by smaller halos. 
This solution, unfortunately, does not solve the problem with the 
NICMOS fields and the narrow $J$-band search. The reason is well exemplified in Fig. 1, where we show that at $z\approx 10$, for the expected Ly$\alpha$ fluxes, NICMOS would be able to detect stellar masses as small as $M_\star = 2\times 10^5 \Msun$, or $M_h \approx 2\times 10^6 \Msun$.
This figure is increased to $M_h \approx 5\times 10^7 \Msun$ for VLT/ISAAC.
Hence chemical feedback cannot provide a way out.                 

We can compute a strong limit on the fraction of NIRB light produced by PopIII stars based on the three candidate $z=10$ galaxies in the UDFs. 

Even if all candidates turn out to be genuine $z=10$ PopIII galaxies, the contribution to the NIRB is less than $0.05 (0.04)$~nW~m$^{-2}$~sr$^{-1}$ in the $J$ ($H$) band, i.e. $\approx 1/340$ of the minimum observed NIRB excess.  
Within these limits, the predicted contribution to {\it Spitzer} counts at larger wavelengths is completely negligible and no sources should be present in the ZEN search.

Alternatively, if PopIII ionizing photons largely escape from star forming 
environments, the Ly$\alpha$ line could be produced in the intergalactic medium. The consequent strong surface brightness decrease would prevent detection and association with the parent galaxy; however, these Ly$\alpha$ photons would still contribute to the NIRB. This scenario offers a possible explanation for the null ZEN result. However, the stellar (as opposed to nebular) emission would make $z\approx 10$ PopIII clusters observable as $J$-dropouts even in this case. The three NICMOS candidates (corresponding to $f_\star^m=1.5$\%) imply then a small NIRB contribution from PopIII, amounting only to $1/24$ of the minimum 
observed excess. Again, these sources would be below detectability for {\it Spitzer} at larger wavelengths.

In conclusion, we find that if the NIRB excess is made entirely by high redshift PopIII clusters, these should have already been observed in deep galaxy searches in large numbers. The few (possibly none!) detections of such objects offers little room for the PopIII explanation of the NIRB excess, leaving us with a puzzling question concerning the origin of this mysterious background component.

\section{Discussion}

Recent measurements of the diffuse background light in the NIR bands have
shown an intensity excess with respect to the light produced by known galaxies
in deep field surveys. Theoretical models have suggested that this 
excess could be produced by redshifted light from the first very massive 
($M>100\;\Msun$) PopIII stars (Santos et al. 2002; SF)         
if these form efficiently down to $z\approx 8$. 
The same stars can also account for the observed small scale angular 
fluctuations detected in the same bands (Magliocchetti et al. 2003).
We have reanalysed this conclusion in the light of new {\it Spitzer} MIR          
observations, and recent high-$z$ galaxy searches.

Our main result is that any model reproducing the minimum value of the excess with emission from PopIII clusters, inevitably predicts that a large
fraction (in some case all) of IRAC (3.5 $\mu$m and 4.5 $\mu$m) counts at the faintest fluxes is due to galaxies at $z>8$. Even more seriously, the same models overpredict the number of $J$-dropouts and Ly$\alpha$ emitters identified in deep field searches by a factor of thousands. In particular, the interpretation of the background light $J$-band peak as due to redshifted  Ly$\alpha$ emission from sources at $z\approx 9$ would imply a large number (394/683 depending on the main gas cooling agent) of identifications in narrow $J$-band surveys. Instead, no such sources have been detected by the ZEN search. 

The three NICMOS $z\approx 10$ candidate PopIII galaxies limit the contribution of such sources to the NIRB excess to be $\le 1/340$. This constraint is somewhat looser if the Ly$\alpha$ emission originates in the IGM rather than in the star forming nebula. In this case the NIRB excess fraction due to PopIII clusters is $\le 1/24$. 

As we are facing such a severe problem, it is wise to consider possible NIRB 
data contaminations from spurious components. The largest experimental 
uncertainty is represented by zodiacal light subtraction. Although it is in 
principle possible to explain the excess in the NIRB with zodiacal light,
this might imply a revision of the properties of dust grains canonically 
assumed (Dwek, Arendt \& Krennich 2005). On the other hand, the existence of the NIR excess above the contribution of deep field galaxies at $z < 5$ 
seems to be suggested by the analysis of the TeV spectral features of distant blazars revealed by Cherenkov experiments
(although different interpretations cannot be excluded, see Aharonian et al. 2005). 
Finally, zodiacal light has been excluded as a possible source of the observed NIRB angular fluctuations 
(Kashlinsky et al. 2002, 2005).  

The interpretation of the NIRB excess must then be different. Are there viable alternatives ? Cooray \& Yoshida (2004) have speculated that a fraction of the unresolved NIRB can be due to high-$z$ miniquasars. However, this hypothesis can be readily discarded by noting that emission from a population of early miniquasars significantly contributing to the NIRB, largely exceeds the unresolved fraction of the Soft X-ray Background (Salvaterra, Haardt \& Ferrara 2005; Madau \& Silk 2005). In particular, Salvaterra et al. (2005) have shown that the SXRB constrains the contribution of these sources
to be $\le 10^{-3}$~nW~m$^{-2}$~sr$^{-1}$.  High redshift supernovae have also be examined by Cooray \& Yoshida (2004), who exclude any significant NIRB contribution from such objects. Finally, more exotic sources as sterile neutrinos have been firmly excluded as viable explanations for the excess by Mapelli \& Ferrara (2005). 

What are the remaining possibilities ? Local, low surface brightness galaxies undetected by available instruments might reduce the gap between NIRB data and the integrated light of deep field galaxies; similarly the light from the outer regions of moderate redshift galaxies could be missed (Bernstein, Freedman \& Madore 2002). Although the nature and properties of such unidentified objects are unknown (note that these sources must have $J >29$), these local galaxies would not explain the excess in the small angular scale fluctuation observed in the NIRB in the $J$ band (Kashlinsky et al. 2002). A more promising alternative would be to locate these faint sources at $z\approx 2-3$, for example if the Lyman Break Galaxies
luminosity function is extremely steep. 

In any case a considerable amount of study will be necessary to fully understand the puzzling nature of the NIRB excess.

\section*{Acknowledgements}
We thank R. Schneider, R.J. Bouwens, E. Komatsu, M. Mapelli, D. Schaerer and
J. Willis for insightful comments.

\end{document}